# Tunable Multifunction Filter Using Current Conveyor

*Manish Kumar*
Electronics and Communication Engineering Department
Jaypee Institute of Information Technology
Noida, India
manishkumar.jiit@gmail.com

*M.C. Srivastava*
Electronics and Communication Engineering Department
Jaypee Institute of Information Technology
Noida, India
mc.srivastava@jiit.ac.in

*Umesh Kumar*
Electrical Engineering Department
Indian Institute of Technology
Delhi, India
drumeshkumar98@rediffmail.com

*Abstract*—The paper presents a current tunable multifunction filter using current conveyor. The proposed circuit can be realized as on chip tunable low pass, high pass, band pass and elliptical notch filter. The circuit employs two current conveyors, one OTA, four resistors and two grounded capacitors, ideal for integration. It has only one output terminal and the number of input terminals may be used. Further, there is no requirement for component matching in the circuit. The resonance frequency ($\omega_0$) and bandwidth ($\omega_0/Q$) enjoy orthogonal tuning. The cutoff frequency of the filter is tunable by changing the bias current, which makes it on chip tunable filter. The circuit is realized by using commercially available current conveyor AD844 and OTA LM13700. A HSPICE simulation of circuit is also studied for the verification of theoretical results.

I. INTRODUCTION

Active filters with current/voltage controllable frequency have a wide range of applications in the signal processing and instrumentation area. Tsividis et. al. employed the realization of on chip MOSFET as voltage controlled resistor [1]. Their contributions and several other research papers may be considered to be motivation for the VLSI industry to make on chip tunable filters [2],[3]. These realizations have small range of variation in the frequency. The OTA-C structure is highly suitable for realizing electronically tunable continuous time filters. A number of voltage mode/current mode OTA-C biquad have been reported in the literature. Multiple-input multiple-output (MIMO), multiple-input single-output (MISO) and single-input multiple-output (SIMO) type circuits have also appeared in the literature. In 1996, Fidler and Sun proposed realization of current mode filter with multiple inputs and two outputs at different nodes using four dual output OTA's and two grounded capacitors [4]. Later, Chang proposed multifunctional biquadratic filters, using three operational transconductance amplifiers and two grounded capacitors [5]. In 2003, Tsukutani et. al proposed current mode biquad with single input and three multiple outputs using three OTAs and current follower (CF) [6].

In the recent years there has been emphasis on implementation of the voltage mode/current mode active filters using second generation current conveyors (CCIIs) which provide simple realization with higher bandwidth, greater linearity and larger dynamic range. Kerwin-Huelsman-Newcomb (KHN) biquad realization of low-pass, band-pass and high-pass filters with single input and three outputs, employing five current conveyor (CCII), two capacitors and six resistors was proposed by Soliman in 1995 [7]. A universal voltage-mode filter proposed by Higasimura et al. employs seven current conveyors, two capacitors and eight resistors [8]. Realization of high-pass, low-pass and band-pass filters using three positive current conveyor and five passive components was reported by Ozoguz et. al.[9]. Chang and Lee [10] and subsequently Toker et. al. [11] proposed realization of low-pass, high-pass and band-pass filters employing current conveyors and passive components with specific requirements. Manish et. al. [12] proposed the realization of multifunction filter (low-pass, high-pass, band-pass and notch filters) with minimum current conveyors and passive components. The central/cutoff frequency of these realizations could be changed by changing the passive components.

In 2001 Wang and Lee implemented insensitive current mode universal biquad MIMO realization using three balanced output current conveyors and two grounded capacitors [13]. In 2004 Tangsrirat and Surakampontorn proposed electronically tunable current mode filters employing five current controlled current conveyors and two grounded capacitors [14]. A tunable current mode multifunction filter was reported in 2008 using five universal current conveyors and eight passive components [15]. Recently Chen and Chu realized universal electronically controlled current mode filter using three multi-output current controlled conveyors and two capacitors, however the frequency and quality factor of their realizations are not independent [16].

The proposed realization in this paper employs two current conveyors, one OTA, five resistors and two grounded capacitors with one output terminal and three input terminals. All the basic low-pass, high-pass, band-pass and notch filters may be realized by the proposed circuit by selecting proper input terminals. The frequency of the filter can be changed by changing the control voltage of the OTA.

The following section presents circuit description of the current conveyor. The sensitivity analysis, simulation results and conclusion are discussed in the subsequent sections.

## II. CIRCUIT DESCRIPTION

The first and second generation current conveyors were introduced by Sedra and Smith in 1968, 1970 respectively; these are symbolically shown in fig 1 and are characterized by the port relations given by "(1)"

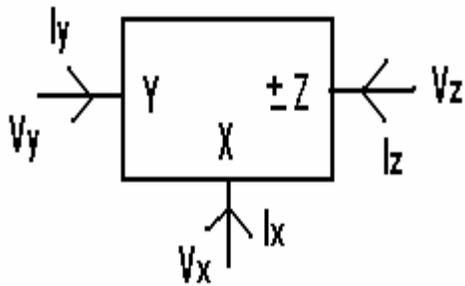

Figure 1. Symbol of Current Conveyor II

$$\begin{bmatrix} V_x \\ I_y \\ I_z \end{bmatrix} = \begin{bmatrix} 0 & B & 0 \\ 0 & 0 & 0 \\ \pm K & 0 & 0 \end{bmatrix} \begin{bmatrix} I_x \\ V_y \\ V_z \end{bmatrix} \quad (1)$$

The values of B and K are frequency dependent and ideally B=1 and K=1. The ±K indicates the nature of current conveyor. +ve sign indicates positive type current conveyor while –ve sign indicates negative.

The proposed circuit shown in fig 2 employs only two current conveyors, five resistors and two capacitors. The grounded capacitors are particularly very attractive for the integrated circuit implementation.

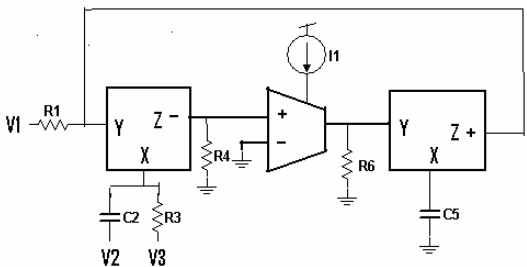

Figure 2. Proposed Voltage Mode Multifunction Filter

The current $I_{abc}$ and parameter $g_m$ may be expressed as follows

$$g_m = \frac{I_{abc}}{2V_T}$$

Where $V_T = KT/q$ is the thermal voltage. The routine analysis yields the following transfer function:

$$V_{out} = \frac{1}{D(s)} \begin{pmatrix} s^2 C_2 C_5 R_1 R_3 R_4 R_6 g_m V_2 + R_3 V_1 + \\ sC_5 g_m R_1 R_4 R_6 V_3 \end{pmatrix} \quad (2)$$

Where

$$D(s) = s^2 C_2 C_5 g_m R_1 R_3 R_4 R_6 + sC_5 g_m R_1 R_4 R_6 + R_3 \quad (3)$$

Thus by using "(2)" we can realize low-pass, band-pass, high-pass and notch filter responses at the single output terminal by applying proper inputs at different nodes as shown in table1.

TABLE I. VARIOUS FILTER RESPONSES

| Filter\Input | $V_1$ | $V_2$ | $V_3$ |
|---|---|---|---|
| Low-pass | 1 | 0 | 0 |
| High-pass | 0 | 1 | 0 |
| Band-pass | 0 | 0 | 1 |
| Notch | 1 | 1 | 0 |

The denominators for the all filter responses are same. The filtering parameter cutoff frequency ($\omega_o$), bandwidth ($\omega_o/Q$) and quality factor (Q) are given by

$$\omega_0 = \sqrt{\frac{1}{R_1 R_4 R_6 C_2 C_5 g_m}} \quad (4)$$

$$\frac{\omega_0}{Q} = \frac{1}{R_3 C_2} \quad (5)$$

$$Q = R_3 \sqrt{\frac{C_2}{g_m R_1 R_4 R_6 C_5}} \quad (6)$$

It can be seen from a perusal of "(4)" - "(6)" that the center frequency and bandwidth $\omega_0/Q$ can be controlled independently through $R_6$ and /or $C_5$ and $R_3$. The transconductance of the OTA is independently tuned by varying the bias current of the OTA.

### III. SENSITIVITY ANALYSIS

The sensitivity analysis of the proposed circuit is presented in terms of the sensitivity of $\omega_0$ and Q with respect to the variation in the passive components as follows:

$$S^{\omega_0}_{C_2,C_5,R_1,R_4,R_6,g_m} = -\frac{1}{2} \quad (7)$$

$$S^{Q}_{R_3} = 1 \quad (8)$$

$$S^{Q}_{g_m,R_1,R_4,R_6,C_5} = -\frac{1}{2} \quad (9)$$

$$S^{Q}_{C_2} = \frac{1}{2} \quad (10)$$

As per these expressions, both the $\omega_0$ and Q sensitivities are less than ± ½ with a maximum value of $S^{Q}_{R_3} = 1$.

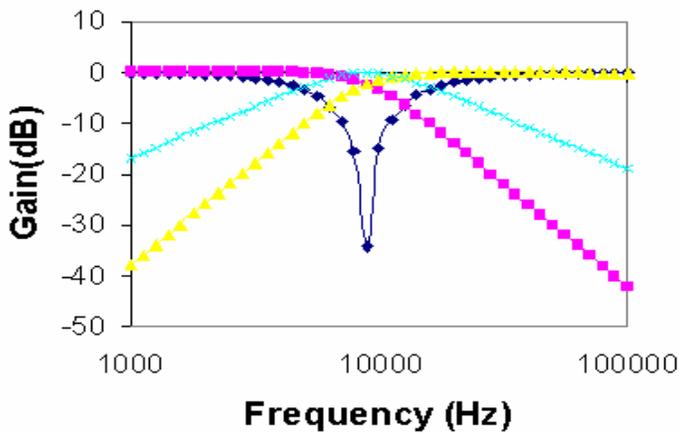

Figure3: Multifunction Filter Response

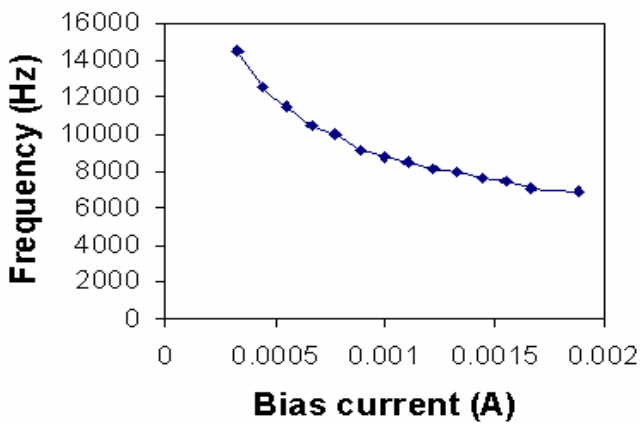

Figure4: Frequency Vs Control Bias Current

### IV. SIMULATION RESULT

The complete circuit is simulated using commercially available AD844 and LM13700. The AD844 is used for the realization of CCII+ and CCII-. Figure 3 displays the simulation result for the proposed filter. The circuit is designed for $\omega_0$ = 8.7 KHz and Q=0.12 by considering $R_1 = R_4 = R_6 = 10K\Omega$, $C_2 = C_5 = 10nF$, $R_3 = 14K\Omega$ and gm=13.2mS. The theoretical results have been are verified to match with simulation result. Figure 3 shows that the quality factor of the notch filter is very high. It is due to the transfer function of the notch filter is having complex conjugate zeros with zero real values. Figure 4 shows the cutoff/center frequency of the filter with respect to the changes in bias current of the OTA. The response is showing the when the bias current is higher than the output current of the OTA than the frequency variation is linear and circuit will be stable.

### V. CONCLUSION

The circuit proposed in this paper generates low-pass, high-pass, band-pass and notch filter using two current conveyors, four resistors and two capacitors. The circuit provides more number of filter realizations at the single output terminal. In addition of this proposed circuit does not have any matching constraint/cancellation condition. The circuit employs' grounded capacitor, suitable for IC fabrication. The circuit enjoys the othogonality between the cutoff frequency and the bandwidth. The OTA is linearly tunable when the bias current is higher than the output current. It has low sensitivities figure of both active and passive components.

REFERENCES

[1] Y. Tsividis, M. Banu and J. Khoury," Continious –time MOSFET-C filters in VLSI," IEEE journal of solid-state circuits, vol. sc-21, no.1, pp. 15-30, 1986.

[2] M. Ismail, S. V. Smith and R. G. Beale, ".A new MOSFET-C universal filter structure for VLSI," IEEE journal of solid-state circuits, vol. sc-23, no.1, pp. 182-194, 1988.

[3] Jaap van der Plas"MOSFET-C Filter with Low Excess Noise and Accurate Automatic Tuning",IEEE Journal of Solid State Circuits, vol. 26, no. 7, pp.922-929, 1991.

[4] Yichuang Sun and J. K. Fidler ,"Structure Generation of Current-Mode Two Integrator Loop Dual Output-OTA Grounded Capacitor Filters," IEEE trans. on cas-II: analog and digital signal processing, vol.43, no.9, pp. 659-663, 1996.

[5] Chun-Ming Chang," New Multifunction OTA-C Biquads," IEEE trans. on cas-II: analog and digital signal processing, vol..46, no.6, pp. 820-824, 1999.

[6] T. Tsukutani, Y. Sumi, M. Higashimura and Y. Fukui," Current-mode biquad using OTAs and CF," Electronics letters, vol. 39, no-3, pp 262-263, 2003.

[7] A. M. Soliman, "Kerwin–Huelsman–Newcomb circuit using current conveyors," *Electron. Lett.,* vol. 30, no. 24, pp. 2019–2020, Nov. 1994.

[8] M. Higasimura and Y. Fukui, "Universal filter using plus-type CCII's," *Electron. Lett*. vol. 32, no. 9, pp. 810-811, Apr. 1996.

[9] S. Ozoguz, A. Toker and O. Cicekoglu, "High output impedance current-mode multifunction filter with minimum number of active and reduced number of passive elements," *Electronics Letters*, vol. 34, no 19, pp. 1807-1809, 1998

[10] Chun-Ming Chang and Ming- Jye Lee, "Voltage mode multifunction filter with single input and three outputs using two compound current conveyors," IEEE Trans. On Circuits and Systems-I: vol. 46, no. 11, pp.1364-1365, 1999.

[11] A. Toker, O. Çiçekoglu, S. Özcan and H. Kuntman ," High-output-impedance transadmittance type continuous-time multifunction filter with minimum active elements," *International Journal of Electronics*, Volume 88, Number 10, pp. 1085-1091, 1 October 2001.

[12] Manish Kumar, M.C. Srivastava and Umesh Kumar," Current conveyor based multifunction filter," International Journal of Computer Science and Information Security, vol. 7 no. 2, pp. 104-107, 2009.

[13] Hung-Yu Wang and Ching-Ting Lee, "Versatile insensitive current-mode universal biquad implementation using current conveyors," IEEE Trans. On Circuits and Systems-I: vol. 48, no. 4, pp.409-413, 2001.

[14] Worapong Tangsrirat and Wanlop Surakampontorn," Electronically tunable current-mode universal filter employing only plus-type current-controlled conveyors and grounded capacitors, "Circuits systems signal processing, circuits systems signal processing, vol. 25, no. 6, pp. 701–713, 2006.

[15] Norbert Herencsar and Kamil Vrba,"Tunable current-mode multifunction filter using universal current conveyors," IEEE third international conference on systems, 2008.

[16] H.P. Chen and P.L.Chu,"Versatile universal electronically tunable current-mode filter using CCCIIs," IEICE Electronics Express, vol. 6, no. 2 pp. 122-128, 2009.

[17] A. M. Soliman, "Current mode universal filters using current conveyors: classification and review," *Circuits syst Signal Process*, vol. 27, pp. 405-427, 2008.

[18] P. V. Anada Mohan, Current Mode VLSI Analog Filters, Birkhauser, Boston, 2003.

## AUTHORS PROFILE

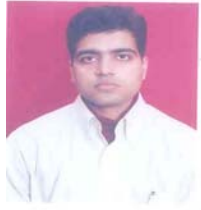

**Manish Kumar** was born in India in 1977. He received his B.E. in electronics engineering from S.R.T.M.U. Nanded in 1999 and M.E. degree from Indian Institute of Science, Bangalore in 2003. . He is perusing Ph.D. He is working as faculty in Electronics and Communication Engineering Department of Jaypee Institute of Information Technology, Noida He is the author of 10 papers published in scientific journals and conference proceedings. His current area of research interests includes analogue circuits, active filters and fuzzy logic.

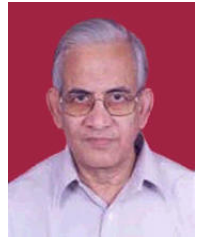

**M**. C. Srivastava  received his B.E. degree from Roorkee University (now IIT Roorkee), M.Tech. from Indian Institute of Technology, Mumbai and Ph.D from Indian Institute of Technology, Delhi in 1974.  He was associated with I.T. BHU, Birla Institute of Technology and Science Pilani, Birla Institute of Technology Ranchi, and ECE Dept. JIIT Sector-62 Noida. He has published about 60 research papers.  His area of research is signal processing and communications. He was awarded with Maghnad Saha Award for his research paper.

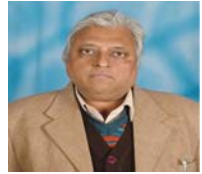

Umesh Kumar is a senior member, IEEE. He received B.Tech and Ph.D degree from IIT Delhi. He has published about 100 research papers in various journals and conferences. He is working as faculty in Electrical Engineering Department, IIT Delhi.